\documentclass[10pt, aps,prl,twocolumn,showkeys,superscriptaddress]{revtex4-2}

\usepackage[english]{babel}             %language support package

\usepackage{placeins}                   % Defines a \FloatBarrier command, beyond which floats may not pass; useful, for example, to ensure all floats for a section appear before the next \section command. 
\usepackage[artemisia]{textgreek}       % Use upright greek letters as text symbols, e.g. \textbeta.
\usepackage{textcomp}
\usepackage{amssymb}   
\usepackage{gensymb}
\usepackage{amsmath} 				    % improve the structure and information in our document with displayed equations
\usepackage{latexsym}                   % additional symbols: \mho  \Join  \Box  \Diamond  \leadsto \sqsubset \sqsupset \lhd  \unlhd  \rhd \unrhd
\usepackage{nicefrac}                   % typesets fractions “nicely” — in the form ‘a/b’ (i.e., staggered with a slash between them, rather than directly one over the other)

\usepackage{graphicx}                   % builds upon the graphics package, providing a key-value interface for optional arguments to the \includegraphics command. This interface provides facilities that go far beyond what the graphics package offers on its own. 
\usepackage{todonotes}
\usepackage{epsfig}                     % AUTODATED: The epsfig package was created as a hybrid of the two graphics packages, with its \epsfig command using the \psfig syntax and much of the more-robust \epsfbox code. Unfortunately, \epsfig still used some of the less-robust \psfig code
\usepackage{epstopdf}                   % a Perl script that converts an EPS file to an ‘encapsulated’ PDF file
\usepackage{dcolumn}                    % makes use of the array package to define a “D” column format for use in tabular environments. 

\usepackage{lipsum}                     % easy access to 150 paragraphs of the Lorem Ipsum dummy text 
\usepackage{color}                      % provides both foreground (text, rules, etc.) and background colour management

\usepackage{subfigure}                  % support for the manipulation and reference of small or ‘sub’ figures and tables within a single figure or table environment. convenient to use when your subfigures are to be separately captioned, referenced, or are to be included in the List-of-Figures.
\usepackage{blindtext}                  % commands \blindtext and \Blindtext for creating ‘blind’ text useful in testing new classes and packages
\usepackage[ulem=normalem]{changes}     % means for manual change markup: \added, \deleted, \replaced
\usepackage{array}                      % extended implementation of the array and tabular environments which extends the options for column formats, and provides “programmable” format specifications
\usepackage{soul}                       % hyphenatable spacing out (letterspacing), underlining, striking out, etc.,
\usepackage[T1]{fontenc}                % select font encodings, and for each encoding provides an interface to ‘font-encoding-specific’ commands for each font
\usepackage[hidelinks]{hyperref}        % handle cross-referencing commands in LaTeX to produce hypertext links in the document

\usepackage[separate-uncertainty=true,multi-part-units=single]{siunitx}

\DeclareSIUnit\sq{\ensuremath{\Box}}
\DeclareSIUnit\bar{bar}
\DeclareSIUnit\angstrom{\text{Å}}
\DeclareSIUnit{\atpercent}{at\%}

\newcolumntype{P}[1]{>{\centering\arraybackslash}p{#1}}
\hypersetup{
    colorlinks,
    linkcolor={red!50!black},
    citecolor={blue!50!black},
    urlcolor={blue!80!black}
}

\newcolumntype{L}[1]{>{\raggedright\arraybackslash}p{#1}}
\newcolumntype{C}[1]{>{\centering\arraybackslash}p{#1}}
\newcolumntype{R}[1]{>{\raggedleft\arraybackslash}p{#1}}

\begin{document}

\title{High-temperature growth of ultra thin NbTiN films on lithium niobate for integrated single photon detection}

%author order to be determined
\author{Sjoerd Telkamp}
\altaffiliation{These authors contributed equally to this work.}
\affiliation{Solid State Physics Laboratory, ETH Zurich, CH-8093 Zurich, Switzerland}
\affiliation{Quantum Center, ETH Zurich, CH-8093 Zurich, Switzerland}
\author{Odiel Hooybergs}
\altaffiliation{These authors contributed equally to this work.}
\affiliation{Solid State Physics Laboratory, ETH Zurich, CH-8093 Zurich, Switzerland}
\affiliation{Optical Nanomaterials Group, Institute for Quantum Electronics, ETH Zurich, CH-8093 Zurich, Switzerland}
\affiliation{Quantum Center, ETH Zurich, CH-8093 Zurich, Switzerland}
\author{Myriam Rihani}
\affiliation{Optical Nanomaterials Group, Institute for Quantum Electronics, ETH Zurich, CH-8093 Zurich, Switzerland}
\affiliation{Quantum Center, ETH Zurich, CH-8093 Zurich, Switzerland}
\author{Giovanni Finco}
\affiliation{Optical Nanomaterials Group, Institute for Quantum Electronics, ETH Zurich, CH-8093 Zurich, Switzerland}
\affiliation{Quantum Center, ETH Zurich, CH-8093 Zurich, Switzerland}
\author{Tummas Napoleon Arge}
\affiliation{Optical Nanomaterials Group, Institute for Quantum Electronics, ETH Zurich, CH-8093 Zurich, Switzerland}
\affiliation{Quantum Center, ETH Zurich, CH-8093 Zurich, Switzerland}
\author{Robin N. D\"urr}
\affiliation{Department of Chemistry and Applied Biosciences, ETH Zurich, CH-8093 Zurich, Switzerland}
\author{Victor Mougel}
\affiliation{Department of Chemistry and Applied Biosciences, ETH Zurich, CH-8093 Zurich, Switzerland}
\author{Daniel Scheffler}
\affiliation{Institute of Physics, Czech Academy of Sciences, 162 00 Prague, Czech Republic}
\author{Filip Krizek}
\affiliation{Institute of Physics, Czech Academy of Sciences, 162 00 Prague, Czech Republic}
\author{Rachel Grange}
\affiliation{Optical Nanomaterials Group, Institute for Quantum Electronics, ETH Zurich, CH-8093 Zurich, Switzerland}
\affiliation{Quantum Center, ETH Zurich, CH-8093 Zurich, Switzerland}
\author{Robert J. Chapman}
\affiliation{Optical Nanomaterials Group, Institute for Quantum Electronics, ETH Zurich, CH-8093 Zurich, Switzerland}
\affiliation{Quantum Center, ETH Zurich, CH-8093 Zurich, Switzerland}
\author{Werner Wegscheider}
\affiliation{Solid State Physics Laboratory, ETH Zurich, CH-8093 Zurich, Switzerland}
\affiliation{Quantum Center, ETH Zurich, CH-8093 Zurich, Switzerland}
%TC:ignore
\begin{abstract}
   Lithium niobate-on-insulator (LNOI) is an emerging photonic platform with high potential for scalable quantum information processing due to its strong second-order nonlinearity. However, little progress has been made in developing on-chip single-photon detectors on LNOI. Niobium titanium nitride (NbTiN) superconducting nanowire single-photon detectors (SNSPDs) are a promising candidate for this application. In this work, we use DC reactive magnetron sputtering to grow high-quality NbTiN thin films using an ultra-high vacuum deposition system with a base pressure lower than $2\times 10^{-10}$ mbar. Enabled by the low concentration of background impurities in this system, we investigate the impact of substrate temperature during NbTiN growth. We achieve four nm thick superconducting films with a critical temperature ($T_{c}$) of 12.3 K grown at a substrate temperature of 825 K. We find that the NbTiN films grow in the (111) orientation and evolve from a porous pillar structure when grown at low temperatures to densely packed fibrous grains at higher temperatures. Furthermore, we demonstrate that the increased substrate temperature reduces the oxygen concentration in our films and improves the overall stoichiometry. In addition, we integrate these films with the LNOI platform and investigate the obtained interface quality. Lastly, we fabricate SNSPDs from the NbTiN film on LNOI and characterize the detector performance.

\end{abstract}
\maketitle

\section{Introduction}

Integrated quantum photonics has promising applications in quantum enhanced computation, communication and sensing \cite{Pelucchi2022}. 
Lithium niobate-on-insulator (LNOI) is a relevant platform for quantum optics experiments due to its strong optical nonlinearity, low optical loss and strong electro-optic response \cite{boes2018,Hu2012}. This allows for the integration of entangled photon-pair sources and reconfigurable circuits using thermo-optic and electro-optic phase shifters \cite{Xin22,Chen22,Wang2018,Yu2022}. To realize a complete monolithic platform for LNOI quantum photonics, the integration of efficient single-photon detectors is required. 
 
State-of-the-art superconducting nanowire single-photon detectors (SNSPDs) are commonly made from NbTiN thin films. They offer high detection efficiencies, low timing jitter, reduced dark counts, short dead times and operation at relatively high temperature \cite{Yang2007,Miki_2009,Holzman2019,Ma2024}. In order to achieve optimal detector performance, the NbTiN layer should have a thickness of around 10 nm or less \cite{Zichi19,luo2023}. This poses a significant challenge in terms of film uniformity and homogeneity. Furthermore, film stoichiometry and crystal structure have been shown to have a crucial impact on SNSPD performance \cite{Zichi19,Jia2015,Holzman2019}. These properties can be controlled by sputtering conditions such as the nitrogen-to-argon ratio \cite{Zhang2024}, the sputtering current \cite{zhang2015}, or the pressure during deposition. Because the superconducting transition temperature is sensitive to changes in crystal quality and stoichiometry, it can be used as a general figure of merit indicating film quality and an early indication for SNSPD performance.  

To date, few experimental implementations of integrated SNSPDs on LNOI photonic platforms have been reported \cite{Lomonte2021,Sayem2020,Colangelo2024}. 
Different approaches have been explored by either depositing the superconducting film after waveguide fabrication \cite{Sayem2020,Colangelo2024} or prior \cite{Lomonte2021}. Depositing the superconductor after waveguide fabrication yielded a relatively high on-chip detection efficiency of $46 \%$, but at the cost of significant waveguide propagation losses. Although these losses can be mitigated by fabrication techniques \cite{Colangelo2024}, this subsequently restrains the superconductor growth and complicates the fabrication process. In contrast, depositing and patterning a NbTiN film prior to waveguide fabrication preserved the low waveguide losses typically achieved for LNOI, but yielded a lower detection efficiency of $27\%$. These results stress the need to study the material platform with a focus on optimizing the growth conditions. 

The main goal of this work is to integrate NbTiN SNSPDs with LNOI and optimize this material platform. In the first part of this study, we optimize the NbTiN film quality by changing the substrate temperature of the magnetron sputtering deposition. This growth parameter has not yet been explored for NbTiN thin films, but we find it to have a strong effect on film quality and superconducting performance. The low oxygen concentration in our sputtering chamber allows us to prevent the increased oxidation that typically occurs in sputtering chambers at elevated temperatures. A maximum $T_{c}$ of 12.3 K is found for NbTiN films of 4 nm thickness for a substrate temperature of 825 K. The effect of substrate temperature on the film topography, chemical composition, and crystallinity is studied by atomic force microscopy (AFM), X-ray photoelectron spectroscopy (XPS), and X-ray diffraction (XRD). 

In the second part of this study we use the developed recipe to deposit the film on an LNOI substrate. The quality of the interface between LNOI and the NbTiN thin film is determined by Scanning Transmission Electron Microscopy (STEM) analysis. Lastly, we show the realization of a single-photon detector on the LNOI substrate. The meandering nanowire devices show high critical current ($J=5.7*10^5 \rm{A}/\rm{cm}^2$), low recovery time ($\tau=2.96 \space \rm{ns}$), low dark counts ($<200$ Hz) and an internal quantum efficiency approaching unity. This integration of efficient single-photon detectors on LNOI is a crucial step towards a fully-integrated photonics platform that would allow for on-chip quantum information processing.

\section{\texorpdfstring{N\lowercase{b}T\lowercase{i}N}{NbTiN} film growth}
The NbTiN thin films in this study are deposited by DC reactive magnetron sputtering in a custom-made UHV chamber with a base pressure lower than $2 \times 10^{-10}$ mBar \cite{Todt2023}. The substrate temperature during deposition was controlled by heating coils directly underneath the substrate and measured with a pyrometer. NbTiN films are grown on two different substrates: 2-inch epi-ready (0001) sapphire wafers and 15x15 mm square cutouts of a 300 nm X-cut LNOI wafer from NanoLN. The substrates were pre-baked at 500 K for 8 hours in the load lock and subsequently outgassed for 60 minutes at growth temperature in the UHV chamber before growth. The pressure during the DC magnetron sputtering was $7 \times 10^{-3}$ mbar, the sputtering current 130 mA and the power around 65 W. After growth, the substrates are kept in UHV at growth temperature for another 60 minutes.

Low temperature DC transport measurements were performed in the Van der Pauw geometry using standard lock-in techniques. Wafers are cleaved by hand directly after growth and contacted using indium. By controlling the temperature and measuring the resistance the critical temperature $T_{\rm c}$ is extracted.

\begin{figure}

    \centering
    \includegraphics[width = 1\linewidth]{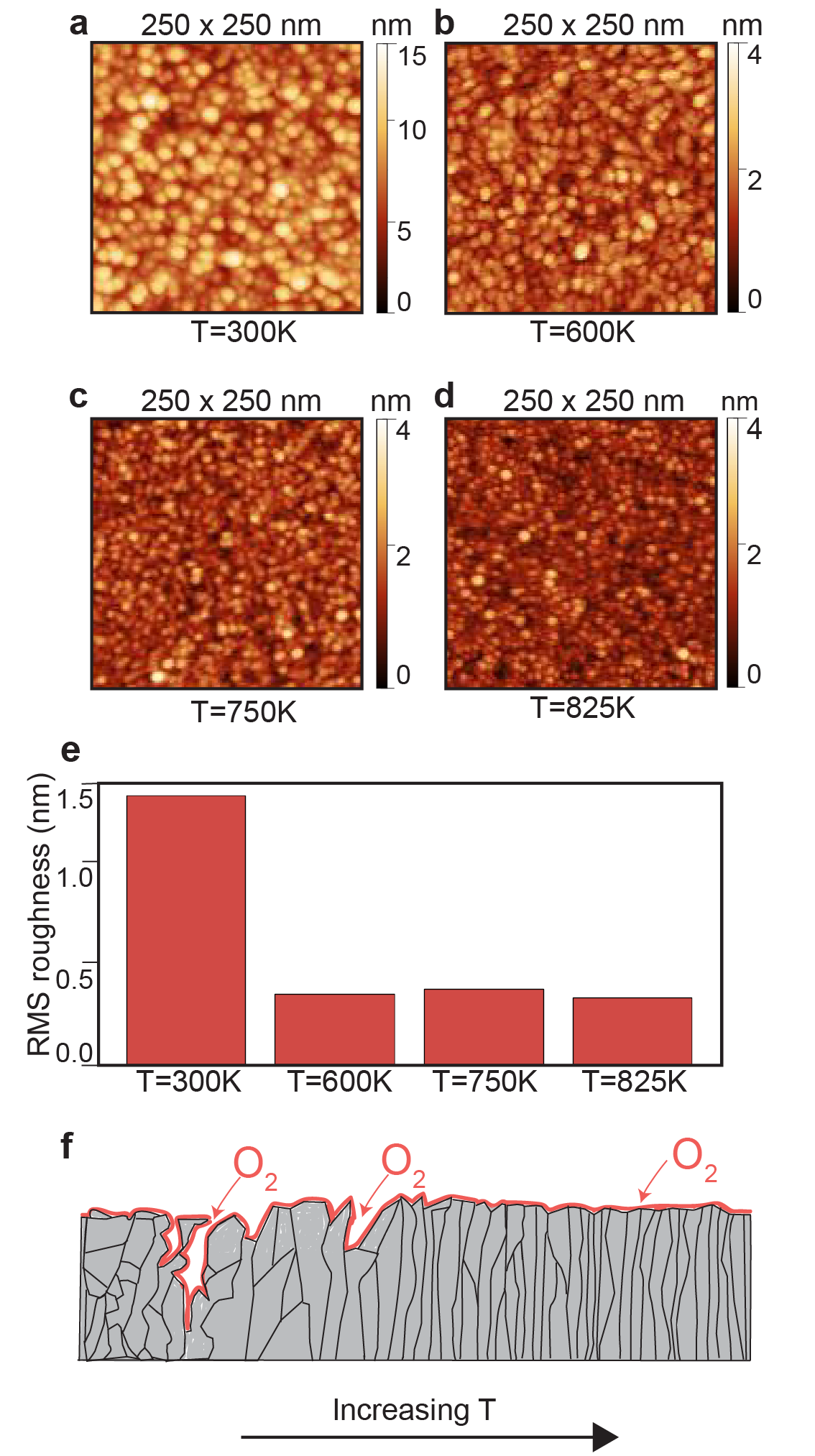}
    
    \caption{\textbf{a} AFM topography showing the obtained NbTiN surface for a substrate temperature of 300 K,  \textbf{b} 600 K,  \textbf{c} 750 K and \textbf{d} 825 K. \textbf{e} The RMS roughness values associated with the various substrate temperatures. \textbf{f} A schematic representation of the expected changes in film structure and how they relate to film oxidation.}
    
    \label{F1}
    
\end{figure}

Figure \ref{F1} shows the surface topography of NbTiN thin films grown at different substrate temperatures measured by AFM. The films have a thickness of 10 nm and are characterized directly after growth. The film grown at room temperature (Fig. \ref{F1}(a)) has a porous and round grain structure typical for polycrystalline NbTiN grown at room temperature on various substrates \cite{Zhang2024}\cite{Steinhauer2020}. Qualitatively, Fig. \ref{F1}(a-d) indicate that an increased growth temperature leads to a more compact columnar structure. This is also reflected by the root mean square (RMS) roughness plotted in Fig. \ref{F1}(e) for a surface area of 1 $\mu$m by 1 $\mu$m, which shows a significant difference between the room temperature film and the films grown at a substrate temperature of 600 K or higher.  

The observed change in surface topography affects the oxygen incorporation in the film, as more oxygen can penetrate into the bulk of the film if the film has a porous structure \cite{halbritter2005transport}. This concept is schematically shown in Fig. \ref{F1}(f), which is based on the Thornton Zone diagram \cite{Thornton1988}. Increased oxygen incorporation will generally degrade the superconducting characteristics such as $T_{\rm{c}}$ and consequently lower SNSPD performance \cite{singh2008,Quintanar_Zamora2024}. Oxidation is typically aggravated by fabrication steps required for patterning the SNSPD, such as heating or plasma ashing. Since the ultra-thin films in this study are especially prone to this effect, we expect the more compact grain structure observed for growth temperatures larger than 600 K to yield better SNSPD performance.

\begin{figure}
 
    \centering
    \includegraphics[width = 1\linewidth]{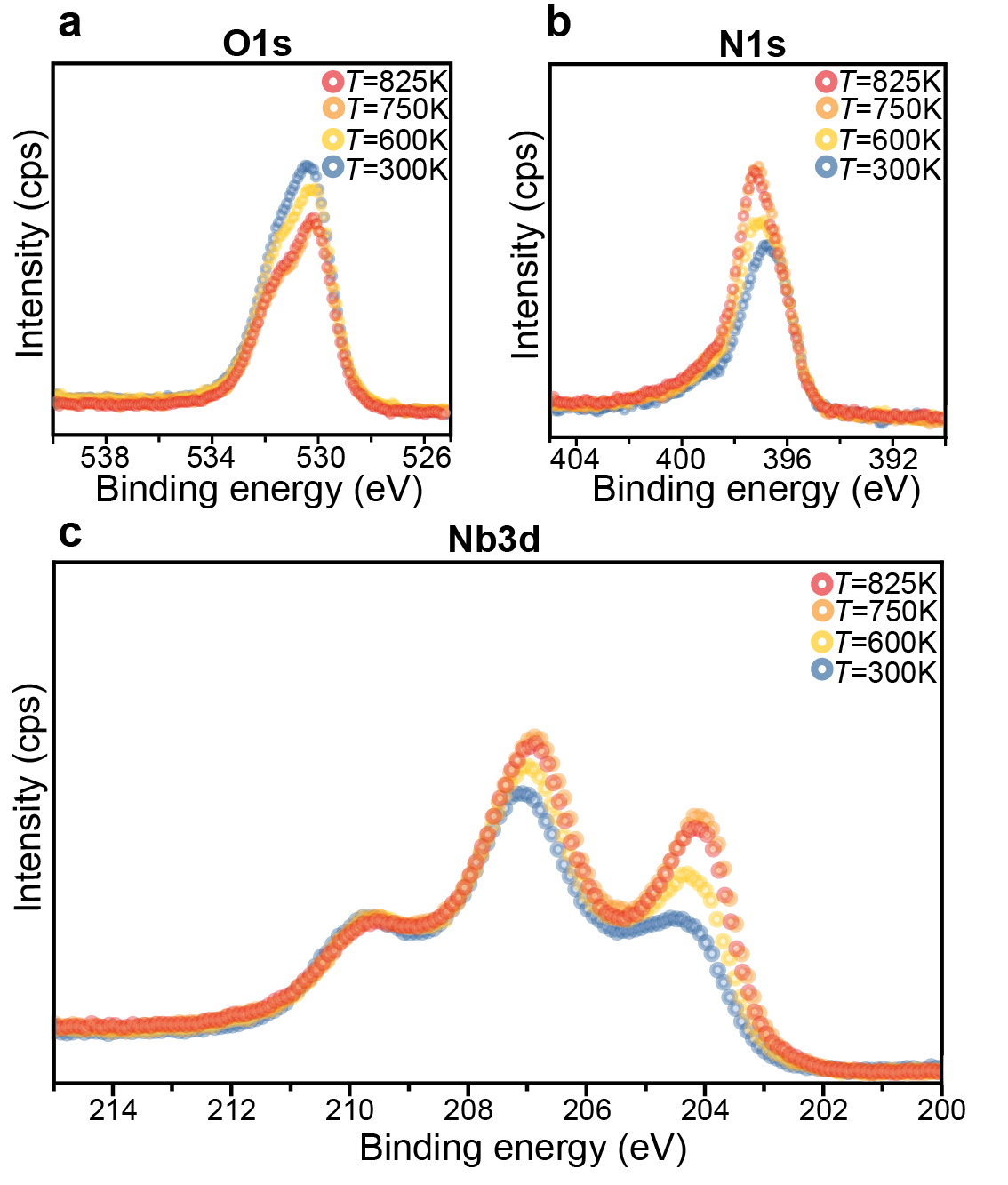}
    
    \caption{\textbf{a} XPS measurements of the NbTiN films grown at various substrate temperatures showing the O1s peak, \textbf{b} N1s peak and \textbf{c} Nb3d peak.}
    
    \label{F2}
    
\end{figure}

The chemical composition of the NbTiN films grown at various substrate temperatures is further studied by XPS analysis shown in Fig. \ref{F2}. Figure \ref{F2}(a) shows that the oxygen content indeed decreases with increasing growth temperature. This decrease saturates for $T >$ 750 K. The O1s peak has two features at 531.5 and 530.0 eV which are associated with oxygen species in the sapphire substrate and metal oxides in the film, respectively \cite{Rotole1998,Rosenberger2008}. 

Similarly, Fig. \ref{F2}(b) demonstrates an increased nitrogen content for films grown at increasingly higher temperature. This is likely associated with thermally increased reactivity during the sputtering process. This increased nitridization saturates for substrate temperatures higher than 750 K. Additionally, a clear shift towards higher binding energy is observed for higher growth temperatures indicating a relatively stronger contribution of metal-nitrides and a reduction of oxidized metal-nitride species \cite{darlinksi1987,milov1995,Prieto1995}. The rising peak at 204 eV in Fig. \ref{F2}(c) indicates a higher concentration of niobium nitride or niobium oxide species. Given the overall decrease in oxygen content seen in Fig. \ref{F2}(a), this is likely due to the NbN contribution. Similar results for the Ti2p peak are shown in the supplementary material \ref{sec_Ti2p}, with a rising contribution of TiN at approximately 455.5 eV \cite{milov1995,Prieto1995}.

\begin{figure}
 
    \centering
    \includegraphics[width = 1\linewidth]{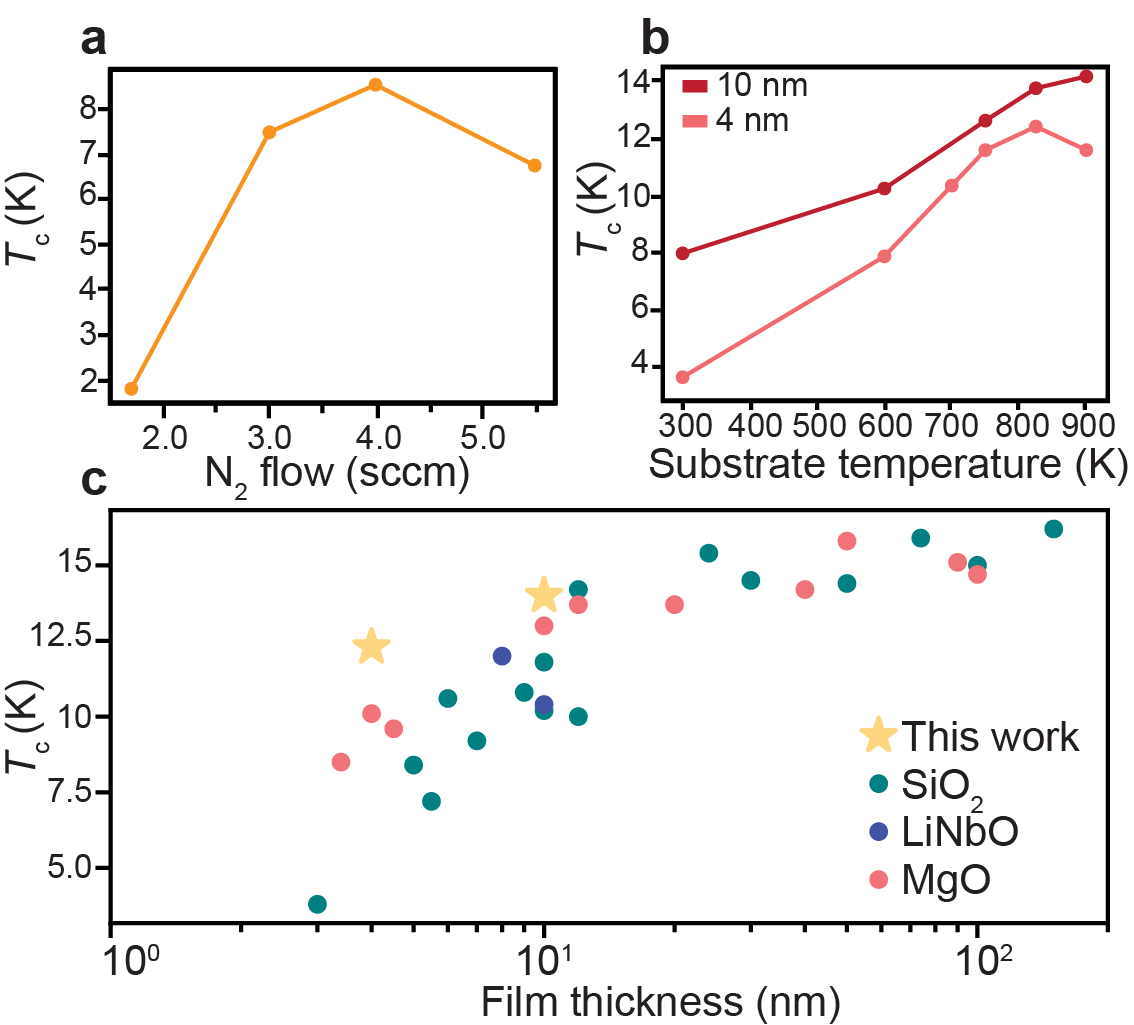}
    
    \caption{ \textbf{a} The $T_{\rm{c}}$ of films grown at increasing N$_2$ flow rate indicating an optimum for 4 sccm. \textbf{b} The $T_{\rm{c}}$ of 4 and 10 nm thick NbTiN films grown at various substrate temperatures at a N$_2$ flow rate of 4 sccm. \textbf{c} The obtained $T_{\rm{c}}$ for our optimal 10 nm film (grown at 900 K) and 4 nm film (grown at 825 K) in comparison to literature results for various substrates. }
    \label{F3}

\end{figure}
 
In Fig. \ref{F3} the superconducting characteristics of NbTiN films grown at various nitrogen flow rates and substrate temperatures are shown and compared to the state-of-the-art. All measurements are done in the van der Pauw geometry using standard DC lock-in techniques in a setup that allows the temperature to be controlled. Figure \ref{F3}(a) shows the effect of nitrogen flow rate on a 10 nm room temperature film. The argon flow rate during the deposition is constant at 50 sccm. An optimum is found for a ratio of 50$/$4 argon-to-nitrogen gas flow,  which agrees well with similar studies \cite{Zhang2024,Steinhauer2020}. %more citations.
This point was chosen as the default nitrogen flow rate for subsequent depositions. 

Figure \ref{F3}(b) shows a significant increase in the superconducting transition temperature for films grown at higher substrate temperature. This is shown for 4 and 10 nm thick films grown on sapphire substrates, with the 10 nm films having generally higher $T_{\rm{c}}$. The increase in $T_{\rm{c}}$ saturates for temperatures higher than 825 K. The respective maximum $T_{\rm{c}}$ values of 12.3 K (at 825 K) and 14.0 K (at 900 K) are, to the best of our knowledge, among the highest reported in literature for films of comparable thickness (Fig. \ref{F3}(c)) \cite{lei2005,Miki_2009,Shiino2010,Guziewicz2013,zhang2015,yang2018,Banerjee_18,Zichi19,Machhadani_2019,Nazir2020,Steinhauer2020,Bretz-Sullivan2022,Bretz-Sullivan2022,Ma_2023,Pratap_2023,gonzalez2023}. This increased superconducting performance is likely related to improved stoichiometry shown by the XPS measurements in Fig. \ref{F2} and reduced oxygen penetration in the bulk of the film, as argued in Fig. \ref{F1}. We associate the observed saturation and even degradation of $T_{\rm{c}}$ for temperatures larger than 825 K to increased oxidation rates during growth. Alternatively, thermally induced stress in the film could also result in the observed degradation of $T_{\rm{c}}$. From these results, we determined the optimum growth temperature to be around 825 K. We find that this optimized growth recipe can be transferred to LNOI substrates, yielding similar $T_{\rm{c}}$ values, which is shown in Appendix \ref{sec_sapphirelnoi}.  

\begin{figure}
 
    \centering
    \includegraphics[width = 1\linewidth]{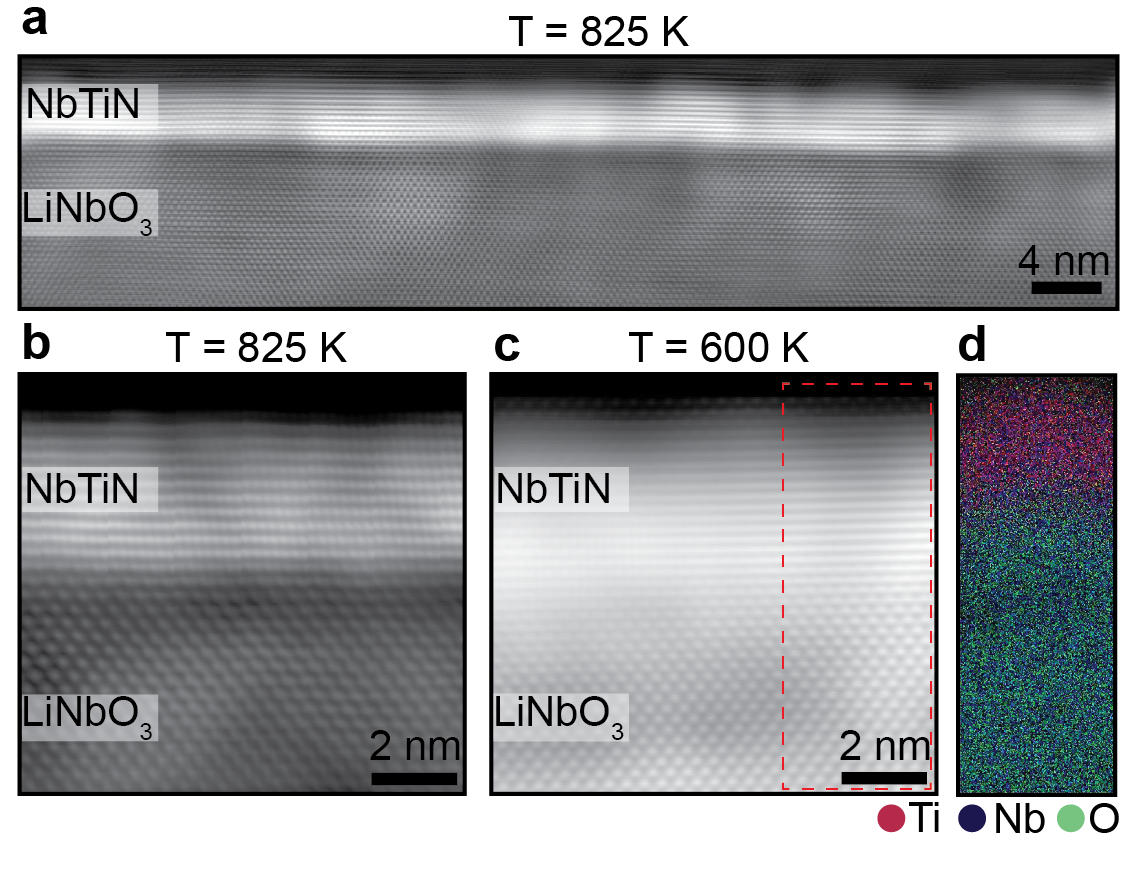}
    
    \caption{\textbf{a} High resolution bright field STEM overview of the 4 nm thick NbTiN film grown at 825 K and the interface with LiNbO$_3$. \textbf{b} Shows the details of the interface between LiNbO$_3$ for $T$ = 825 K \textbf{c} and $T$ = 600 K. \textbf{d} An EDX measurement of the region indicated by the dotted dashed line in \textbf{c}.   }
    
    \label{F4}

\end{figure}

The STEM images in Fig. \ref{F4} show the interface between NbTiN and LiNbO$_3$ for a film grown at 825 K (Fig. \ref{F4}(a) and (b)) and 600 K (Fig. \ref{F4}(c)) on LNOI wafers. A high-quality and sharp interface is observed for both temperatures without clear signs of chemical intermixing. The absence of chemical intermixing is further substantiated by the Energy Dispersive X-ray (EDX) spectroscopy shown in Fig. \ref{F4}(c). The NbTiN film is fully crystalline over the studied 1.5 $\mu$m lamella width and grain boundaries are separated by more than 30 nm. Furthermore, the films are uniform, have no visible thickness variation, and have limited oxide growth. The crystal lattice of the NbTiN film is tightly spaced in the horizontal direction and has clear lattice planes in the vertical direction. This suggests a side projection of the (111) orientation in the growth direction. 

\begin{figure}
 
    \centering
    \includegraphics[width = 1\linewidth]{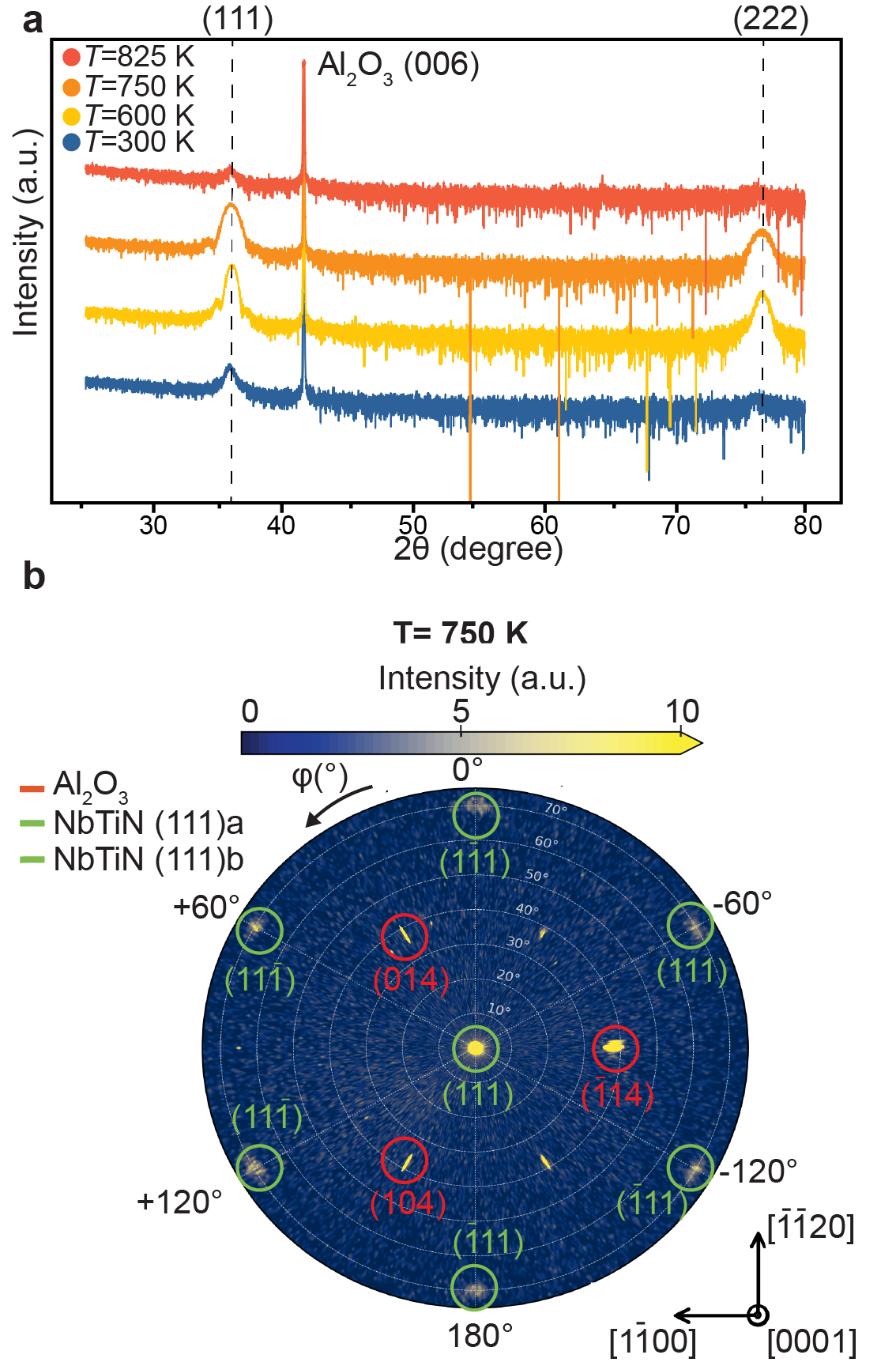}
    
    \caption{\textbf{a} Radial XRD scans of the NbTiN films grown at various temperatures. The only out-of-plane crystal orientation observed over the measured range is the [111] orientation.    \textbf{b} Polar plot of the NbTiN film grown at 750 K with the (111) peak as the central contribution. The six different intensity spots circled in green indicate the two different observed in-plane orientations of the (111) grains. }
    
    \label{F5}
 
\end{figure}

The (111) dominant crystal orientation of the NbTiN films is confirmed in the XRD analysis in Fig. \ref{F5}. In Fig. \ref{F5}(a) radial XRD scans of four different films grown on sapphire substrates at various temperatures are shown. The only observed out-of-plane crystal orientation is (111) and the higher order reflections of that lattice plane. The films grown at 600 K and 750 K show the strongest (111) oriented peaks, indicating the higher crystal quality of these films. This is corroborated by the visible Laue oscillations for these temperatures. Figure \ref{F5}(b) shows a pole figure of the (111) Brags peak from the $T =$ 750 K film in order to further investigate the in-plane orientations of the crystal domains. We observe six different intensity spots under 60 $\deg$, indicating two different in-plane orientations of the (111) crystal domains. 

Overall, a distinct difference in crystal quality can be observed between the room-temperature film and the films with increased substrate temperature. The decrease in oxygen concentration seen in Fig. \ref{F3} combined with the structural changes in the NbTiN film seen in Fig. \ref{F1} and \ref{F5} are the likely root cause for the increased $T_{\rm{c}}$ \cite{yeram2023}. Since we have not observed any intermixing or damaging of the LNOI substrate in the STEM images in Fig. \ref{F4}, we believe the films grown at 825 K will ultimately yield the best detector performance.

\section{SNSPD integration}
Now that the growth of NbTiN films on LNOI has been optimized, we demonstrate the use of this material platform for single-photon detection. Using the growth temperature of $T=825$ K a 10 nm film is grown on LNOI, similar to the films shown in Fig. \ref{F3}. SNSPDs are fabricated by patterning a meandering nanowire structure using electron beam lithography. Etching of the NbTiN is done by reactive ion etching using a mixture of Argon and SF$_6$. More details on the fabrication process are given in appendix \ref{SNSPD_fab}. The fabricated SNSPD device is designed to be a nanowire meandering over a $10$ μm $\times$ $10$ μm area with a duty cycle of 50 $\%$. The nanowire has a width of 100 nm. The device is contacted with a movable radio frequency probe and biased using a current source. All measurements are performed in a cryostat at a temperature of 5 K or higher. 

\begin{figure}
 
    \centering
    \includegraphics[width = 1\linewidth]{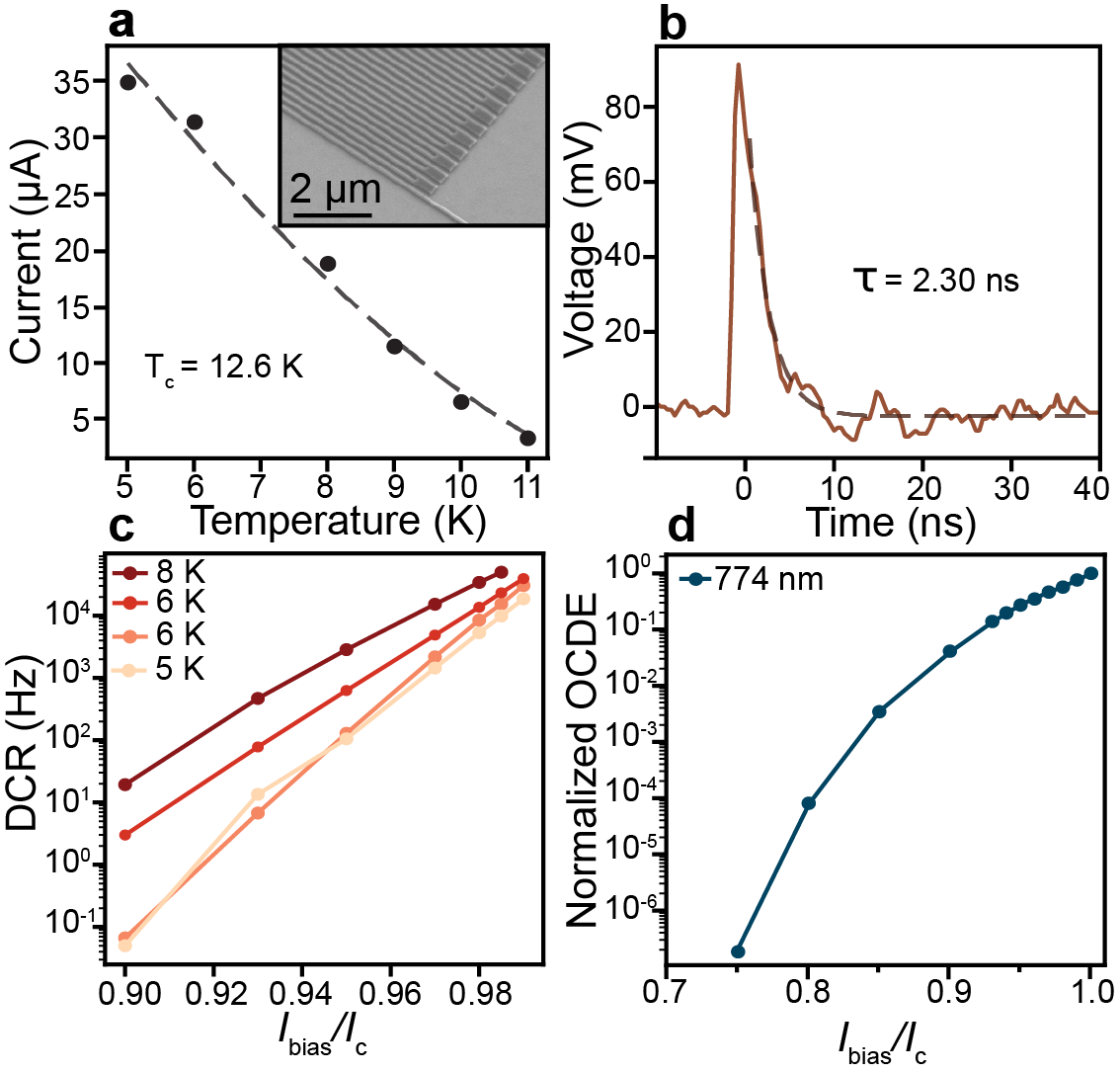}
    
    \caption{\textbf{a} The critical current of the SNSPD as a function of temperature. The dashed line is a fit of the expected critical current relation  with the found fit value of $T_{\rm{c}}$. The inset of the figure shows a scanning electron microscopy image of the device. \textbf{b} The device voltage response after excitation with a single photon. The response time $\tau$ is obtained from the dashed fit and indicated in the figure. \textbf{c} DCR measurements of the detector as a function of normalized current bias for various temperatures. \textbf{d} The normalized detection efficiency as a function of normalized current bias for two different excitation wavelength.}
    
    \label{F6}

\end{figure}

Figure \ref{F6}(a) shows the critical current of the nanowire device plotted as a function of temperature. The inset of Fig. \ref{F6}(a) shows a scanning electron microscopy image of the fabricated SNSPD. The critical current of the wire is fitted using $I_c(T)=I_c(0)\left[1-\left(\frac{T}{T_c}\right)^2\right]^{3 / 2}$ with $T_{\rm{c}}$ as a fitting parameter, and the resulting fit is shown in the dashed line. The obtained $T_{\rm{c}}$ of 12.6 K is relatively close to the $T_{\rm{c}}$ of 13.5 K of the NbTiN film before fabrication, indicating limited oxidation during fabrication. The critical current of the device is relatively large for a wire of this dimension ($J=8.0\times10^5 \rm{A}/\rm{cm}^2$) \cite{Gourgues19,Steinhauer2020,Ma_2023}.  %cite papers. 

Next, we irradiate this device with 774 nm photons using a tunable wavelength laser source. The detection kinetics for operation at a normalized bias of 0.9 are shown in Fig. \ref{F6}(b). The voltage response of the device can be fitted by a single exponential function (dashed line in Fig. \ref{F6}(b)) using a relaxation time $\tau$ of 2.30 ns. This value agrees well with other NbTiN SNSPDs and is significantly shorter than typically found for NbN \cite{Steinhauer2020,Zhang_2020,Dong2024}. This is directly related to the low kinetic inductance of the material. 

Figure \ref{F6}(c) shows the measured dark count rates as a function of bias current for various temperatures. The various temperature traces converge in a single point around $I_{\rm{bias}}=I_{\rm{c}}$. For the lowest temperature trace ($T=5 \space \rm{K}$) the dark counts indicate a reasonable operating regime at a normalized bias current between 0.9 and 0.94 with dark counts below 10 Hz. This is further explored by the normalized on-chip detection efficiency (OCDE) shown in Fig.\ref{F6}(d) as a function of normalized current bias. For a photon wavelength of 774 nm, the detection efficiency starts to saturate around a normalized current bias of 0.9. This saturation indicates that the internal quantum efficiency of the detector is reaching unity. This saturation would likely be reached for lower current bias if the operating temperature was lower \cite{Gourgues19}. We note that the absolute detection efficiencies of these devices are very low ($<1\%$), which we attribute to the device geometry, optical excitation, and absence of an optical cavity which are used in high-efficiency detectors. Future NbTiN hairpin detectors on top of an LNOI waveguide could increase the absolute detector efficiency by enabling a longer interaction time and will enable fully integrated quantum photonics experiments.

In general, the SNSPD device shows performance characteristics similar to those of NbTiN detectors made on more conventional substrates such as Si$\rm{O}_2$ \cite{Miki17,Steinhauer2020,Chang2021}. Based on these results, the optimal operating regime for single-photon detection would be around a normalized bias current of 0.9 and the device can be operated at around 5 K, which is well above typical base temperatures of Gifford-McMahon cryocoolers.  

\section{Conclusion}
In summary, we have shown that increasing the substrate temperature during DC reactive magnetron sputtering of NbTiN can both structurally and chemically improve the quality of the superconducting film. AFM topography and XRD characterization indicate that substrate temperatures exceeding 650 K result in compact films with a columnar structure consisting of (111) oriented grains. This results in a maximum $T_{\rm{c}}$ of around 12 K for ultra-thin films of 4 nm thickness, which is among the highest reported for films of such thickness. Furthermore, we demonstrate by STEM analysis that these high-temperature NbTiN films can also be grown on LNOI substrates while maintaining good interface quality. 

The developed material platform was used to fabricate an SNSPD. Low-temperature optical characterization of this device indicates a high critical current density, low response time, low dark counts and an internal quantum efficiency approaching 1. The ideal operating point of the device would be at a normalized current bias of around 0.9. These results demonstrate that the developed material platform can be used for single-photon detection. This represents a critical step towards fully integrated on-chip photonics on LNOI. In future work, single-photon sources and waveguides could be integrated on the same chip. The on-chip integration of the single-photon detector will allow for faster detection and lower losses, which could enable more advanced photonic circuits involving boson sampling or feed-forward operations. 
% \newpage

\section{Acknowledgment}
We acknowledge financial support from the Swiss National Science Foundation (SNSF) and the NCCR QSIT (National Center of Competence in Research - Quantum Science and Technology). R.J.C. acknowledges support from the Swiss National Science Foundation under the Ambizione Fellowship Program (project number 208707). R.J.C. and W.W. acknowledge support from an ETH Research Grant. R.G. acknowledges support from the Swiss National Science Foundation under the Bridge Program (Project Number 194693). We also acknowledge financial support by the Czech Science Foundation (Grant No. 22-22000M), MEYS grant LM2023051 and  Lumina Quaeruntur fellowship LQ100102201 of the Czech Academy of Sciences.

\bibliography{papersAPS.bib}
\appendix
\section{XPS analysis}

X-ray photoelectron spectroscopy (XPS) was conducted using a Sigma 2 instrument from Thermo Fisher Scientific using a non-monochromatic 200 W Al Kα source, an Alpha 110 hemispherical analyzer, and a seven-channel electron multiplier. Charge corrections were applied by aligning the C 1s peak of adventitious carbon to 284.8 eV for all sample spectra. High-resolution spectra were collected with a pass energy of 20 eV, a step size of 0.1 eV, and a dwell time of 50 ms. All XPS data were analyzed using CasaXPS software \cite{FAIRLEY2021100112}.

\begin{figure}
 
    \centering
    \includegraphics[width = 1\linewidth]{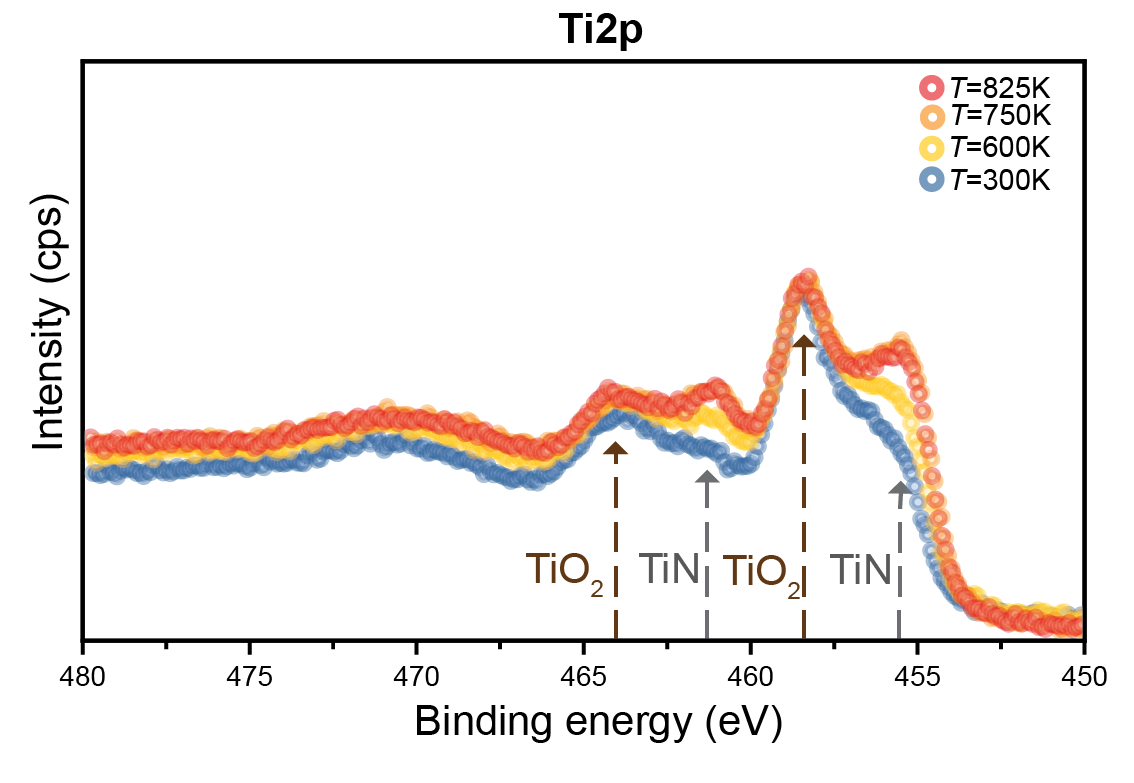}
    
    \caption{\textbf{XPS analysis of the Ti2p peak} \textbf{(a)} XPS analysis of the Ti2p that shows contributions of the TiN ( gray) and the TiO$_2$ (brown) in the various films. }
    
    \label{sup_Ti}
    
\end{figure}

\label{sec_Ti2p}
Figure \ref{sup_Ti} shows XPS analysis of the Ti2P for the 4 films discussed in the main text. The contributions of the TiN and TiO$_2$ are indicated by the gray and brown arrows respectively. We interpret these results as indicating that the contribution of metal nitrides increases with temperature. This further substantiates the overall trend seen in the XPS data of increasing nitrogen content with increasing substrate temperature. 
\section{NbTiN film quality comparison for Sapphire and LNOI substrates}
\label{sec_sapphirelnoi}
\begin{figure}
 
    \centering
    \includegraphics[width = 1\linewidth]{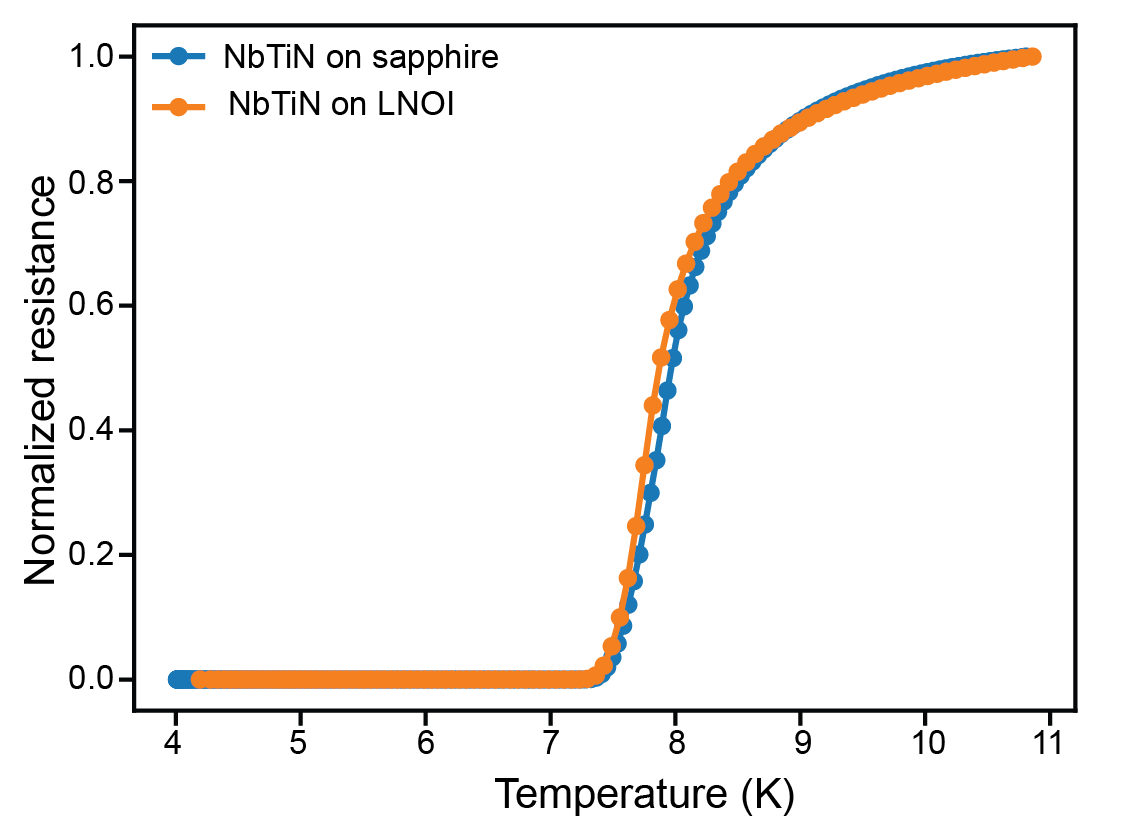}
    
    \caption{\textbf{A comparison between two identical 4 nm NbTiN films deposited on sapphire and LNOI.}}
    
    \label{Tc_compare}

\end{figure}

Figure \ref{Tc_compare} shows a comparison of two identical 4 nm thick NbTiN films that where grown on a LNOI and sapphire substrate for comparison. Both films are grown at a substrate temperature of 600 K. The transition temperature for both superconducting films is 7.8 K, with the differences in the measurement accuracy range. This indicates that the quality of the NbTiN films grown on sapphire and LNOI is similar \cite{Steinhauer2020}. 

\section{SNSPD fabrication}
\label{SNSPD_fab}
Single photon detectors are fabricated using electron beam lithography (EBL), on 100 nm hydrogen silsesquioxane (HSQ) resist, and reactive ion etching based on sulphur hexafluoride (SF$_6$) chemistry. First, contact electrodes are deposited on the un-patterned NbTiN film using EBL and a lift-off process of 100 nm of gold, with 5 nm of chromium as an adhesion layer. Then, meandering structures are patterned with EBL and etched down using SF6. We avoid introducing Ar in the RIE chemistry to prevent sputtering of the underlying lithium niobate film. The remaining etch mask is left as a protection of the patterned structures, as it is chemically and optically equivalent to silicon dioxide.

\section{SNSPD measurement}

\begin{figure}
 
    \centering
    \includegraphics[width = 1\linewidth]{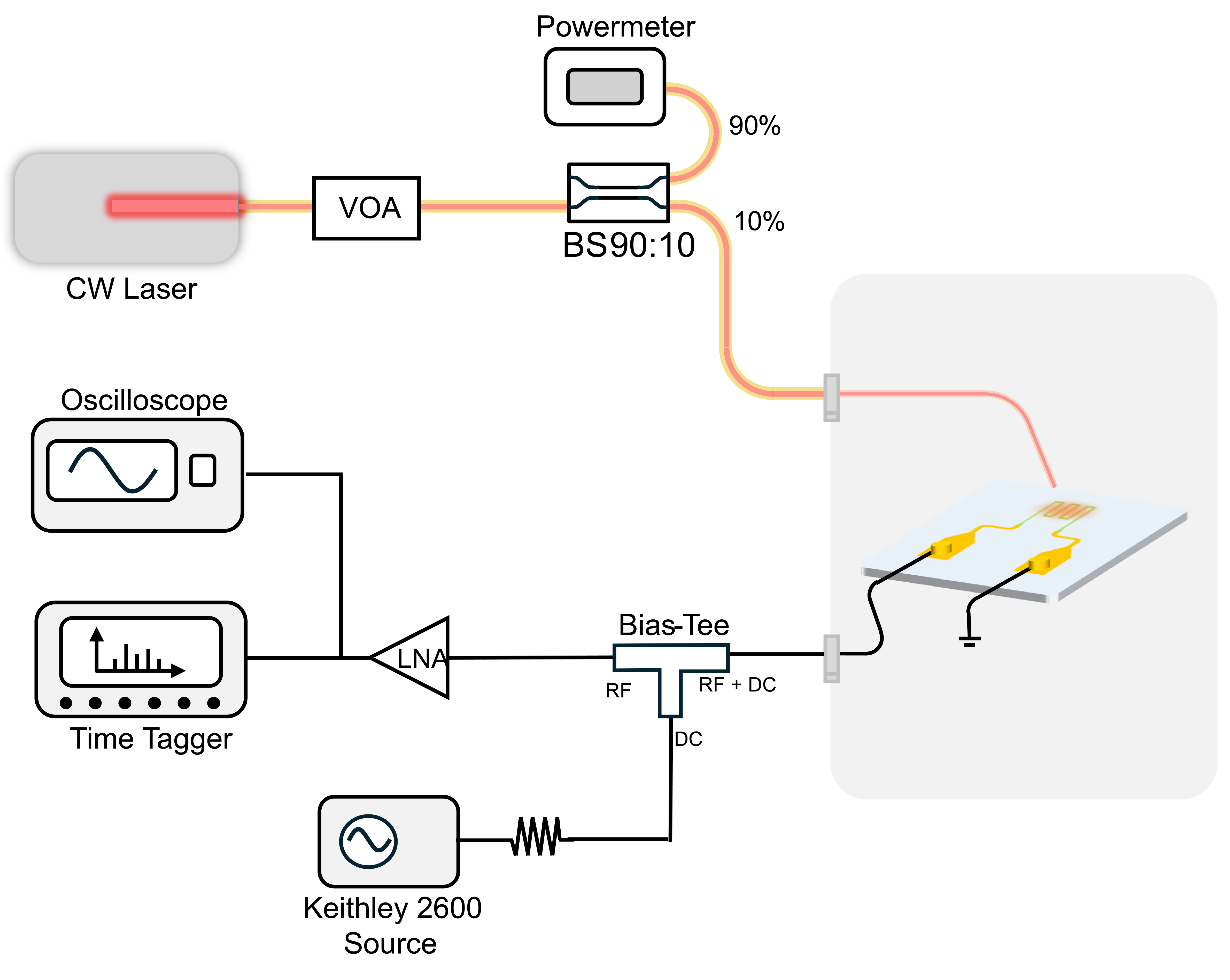}
    
    \caption{An overview of the measurement apparatus used for the SNSPD measurements presented in the main text.   }
    \label{supp_snspd}

\end{figure}

The SNSPD structures are biased using a source-measurement unit (Keithely 2612B) in voltage-source mode in series with a 100kOhm resistor. The source-meter can simultaneously deliver $\mu$A-range bias current and precisely measure the current and resistances across the system with 10pA accuracy. The generated bias signal is then routed to the SNSPD through the DC port of a bias-tee (Z3BT-2R15G+), which helps in blocking noisy RF components. The RF port of the bias tee delivers the voltage pulse of the SNSPD through a Low-Noise Amplifier (AT-LNA-0043-3504Y) that amplifies the signal by 35dB. The amplified signal is then measured by an oscilloscope (Tektronix MDO3052) or time tagger (Swabian Time Tagger Ultra) for counting purposes. The trigger setting of the time tagger must be carefully chosen to filter out any counting errors due to electronic noise, this can be done after inspection of the voltage pulse peak amplitude with the oscilloscope.
For optical excitation we use a continuous-wave tunable laser Toptica CTL-780 for a wavelength of 774 nm. The light is coupled into a variable fiber attenuator (Thorlabs VOA50PM-APC or VOA780PM-APC) then split using a single mode fused fiber 90:10 beam splitter with $\pm$15 nm bandwidth (Thorlabs TN1550R2A2 or TN785R2A1). The 90\% output port is routed to an optical power-sensor (Thorlabs InGaAs Photodiode S154C and Si Photodiode S1501C) to monitor the attenuated optical power while the 10\% output port is coupled to the signal mode fibers inside of the cryostat. The fiber is then aligned on top of the meander structure for optical illumination.
The dark counts rate has been measured using the time tagger at different applied bias averaged over 60 s for each data point. For the on-chip detection efficiency (OCDE), the average number of counts also averaged over 60s. The OCDE is calculated as:
OCDE = (Photon Count Rate - Dark Count Rate) / Photon Flux.
The photon flux reaching the device is estimated by the optical overlap between the fiber output mode and the device structure by comparing the effective area of the nanowire with the fiber mode spot size at the device plane. The fiber mode radius at the device is given by:
$r = (d_{\rm{fiber}} / 2) + L * \tan(\arcsin(NA))$
where $d_{\rm{fiber}}$ is the fiber core diameter, $L$ is the fiber–device distance, and $NA$ is the fiber numerical aperture. The corresponding optical mode surface is:
$A_{\rm{mode}} =\pi r^2$.
Considering $A_{\rm{NW}}$ as the nanowire area, the optical overlap factor is then expressed as:
$\eta = A_{\rm{NW}} / A_{\rm{mode}}$. 

\end{document}